\documentclass[fleqn,usenatbib]{mnras}

\usepackage{newtxtext,newtxmath}
\usepackage{tablefootnote}

\usepackage[T1]{fontenc}
\usepackage{ulem}

\DeclareRobustCommand{\VAN}[3]{#2}
\let\VANthebibliography\thebibliography
\def\thebibliography{\DeclareRobustCommand{\VAN}[3]{##3}\VANthebibliography}

\usepackage{graphicx}	
\usepackage{subcaption}

\def \scbe {SXP 31.0} 
\def \bexrb {BeXRB}
\def \bexrbs {BeXRBs}

\def \swift {Swift}

\title[SXP 31.0]{SXP 31.0  - the 2025 near-Eddington double X-ray outburst after 26 years of quiescence.}

\author[M. J. Coe et al.]
{M. ~J. Coe,$^{1}$\thanks{E-mail: mjcoe@soton.ac.uk}
T.~M. Gaudin$^{2}$,
I.~M. Monageng$^{3,4}$,
J.~A. Kennea$^{2}$,
D.~A.~H. Buckley$^{3,4}$,
  \newauthor
A. Udalski$^{5}$,
P.~A. Evans$^{6}$,
S.~Chattopadhyay.$^{3,7}$\\
$^{1}$Physics \& Astronomy, The University of Southampton, Southampton, SO17 1BJ, UK\\
$^{2}$Department of Astronomy and Astrophysics, The Pennsylvania State University, 525 Davey Lab, University Park, PA 16802, USA\\
$^{3}$South African Astronomical Observatory, P.O Box 9, Observatory, 7935, Cape Town, South Africa\\
$^{4}$Department of Astronomy, University of Cape Town, Private Bag X3, 7701 Rondebosch, South Africa\\
$^{5}$Astronomical Observatory, University of Warsaw, Al. Ujazdowskie 4, 00-478 Warszawa, Poland \\
$^{6}$University of Leicester, X-ray and Observational Astronomy Research Group, School of Physics \& Astronomy, University Road, Leicester LE1 7RH, UK
\\
$^{7}$Centre for Space Research, North-West University, Potchefstroom Campus
Private Bag X6001, Potchefstroom, 2520, South Africa
}
\date{}

\pubyear{2025}

\begin{document}
\label{firstpage}
\pagerange{\pageref{firstpage}--\pageref{lastpage}}
\maketitle

\begin{abstract}

\scbe{} is an X-ray source in the Small Magellanic Cloud (SMC) that was first identified as a Be X-ray Binary ( \bexrb) ~system when it went into X-ray outbusrst in 1998. It is now known to consist of an OBe main sequence star and a neutron star with a spin period of 31s. In 2025 a new X-ray outburst phase began with the source exhibiting a luminosities approaching the Eddington limit of $10^{38}$ erg/s.  Unusually, H$\alpha$ images show it has a surrounding halo whose nature has not been clear. In this paper, we report new observations of this halo, including the first multi-fibre Integrated Flux Unit (IFU) observations, which identify this emission as probably a coincidental HII region. The X-ray, UV \& optical data cover a period of $\sim$200d and reveal that the source underwent two bright, back-to-back, Type II outbursts in 2025 - a rare occurrence for any \bexrb ~system.

\end{abstract}




\section{Introduction }

\bexrbs\ are a large sub-group of the well-established category of High Mass X-ray Binaries (HMXB) characterised by being a binary system consisting of a massive mass donor star, normally an OBe type, and an accreting compact object, normally a neutron star. There are several systems in the Magellanic Clouds where the compact object is identified as a white dwarf \citep{gaudin2024, kennea2021,coe2020}. In addition there is one known galactic system, MWC 656, where the accretor has been proposed to be a black hole, but this remains to be confirmed \citep{2014casares,janssens2023J,dzib2025}. 

The Small Magellanic Cloud (SMC) has been known for quite a while now to contain the largest known collection of \bexrbs\ - see, for example, \cite{coe2015, hs2016}. As a result of many observational studies the complex interactions between the two stars continue to produce unexpected surprises. In particular, the unpredictable behaviour of the mass donor OB-type star is major driver in the observed characteristics of such systems, and as a direct result of the rate of mass transfer onto the neutron star systems long-term spin up or spin down changes are observed \citep{klus2014}. 

The study of this large population of BeXRBs is the primary motivation for the Swift SMC Survey (S-CUBED; \citealt{kennea2018}). S-CUBED is a weekly survey, ongoing since 2016, that aims to both identify new BeXRBs and track transient outbursts from known systems This survey utilizes the X-ray Telescope (XRT; \citealt{Roming05}) and UV/Optical Telescope (UVOT; \citealt{burrows05}) of the Swift Observatory. Tiled observations of 142 overlapping tiles for 60s each are obtained in order to obtain spatially-continuous observations of the entire SMC. Each tile is observed with XRT in Photon Counting (PC) mode and UVOT observing the \textit{uvw1}-band. S-CUBED data has been used to identify several new BeXRBs \citep{Kennea2020, Monageng2019, 2024Gaudin} and to observe notable outbursts from known systems (e.g. SMC X-3 \citealt{Townsend2017} and SMC X-2 \citealt{coe2024} ).

In this paper, we report on the recent near super-Eddington outbursts of \scbe ~(also known as XTE J0111.2-7317) .
This X-ray transient \scbe ~was originally discovered
by the Proportional Counter Array (PCA), on the Rossi X-ray
Timing Explorer (RXTE) X-ray observatory, on 29 October
1998 \citep{chak1998a,chak1998b}. The outburst continued until 1999 January and was detected during its decline at the start of the RXTE monitoring campaign of the SMC \citep{laycock2005}. No further detections of \scbe{} ~were reported from this RXTE/SMC campaign which lasted until 2012. A report of QPO behaviour seen in the 2025 outburst has been presented by \cite{salganik2025a, salganik2005b}. These same authors report on the steady spin up of the pulsar during the Type II outburst.

\begin{figure*}
	\includegraphics[width=16cm,angle=-0]{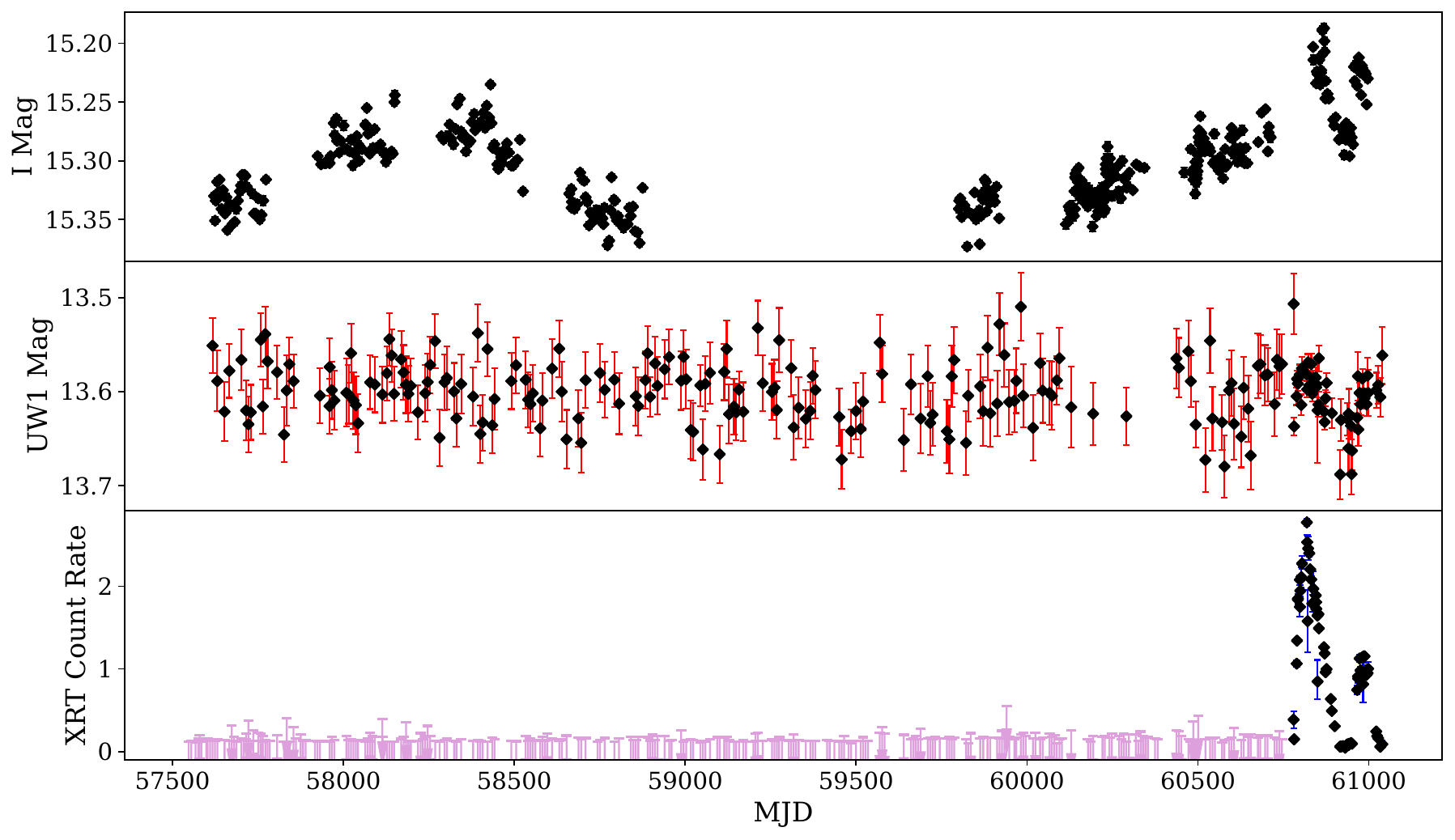}
    \caption{{Full OGLE IV and Swift light curves for \scbe{} spanning the $\sim$9 years duration of the S-CUBED survey.}}
    \label{fig:all_swift}
\end{figure*}

The optical counterpart was identified by \cite{israel1999} and classified by \cite{covino2001} as B0.5V - B1Ve. Subsequently, optical and IR measurements by
\cite{coe2000,coe2003}
confirmed the optical counterpart to be a B type star with a strong IR excess. Unusually,
the H$\alpha$ images of the field reveal an
extended region of emission surrounding the source. \cite{coe2003}
speculated as to whether this very unusual feature was a supernova remnant, a bow shock, or an ionised HII region. These options are re-visited here using higher quality data than was previously available.

The new observations reported here cover the extent of a probable double Type II outburst - each typically with a duration comparable to, or greater than the binary period deduced from our optical observations. In addition each outburst has a luminosity approaching the Eddington limit of $10^{38}$ erg/s. The multi-wavelength observations of the source presented here were obtained by combining Swift XRT and UVOT observations, optical data from the Optical Gravitational Lensing Experiment (OGLE), and contemporaneous imaging and spectroscopic observations from the Southern African Large Telescope (SALT). They reveal a source that had been quiescent for over 26 years exhibiting an extremely bright period of activity.

\section{Observations}

\subsection{\swift{} Observations}

\subsubsection{XRT Observations}
\label{subsubsec:swift_xrt}

\begin{figure*}
	\includegraphics[width=13cm,angle=-0]{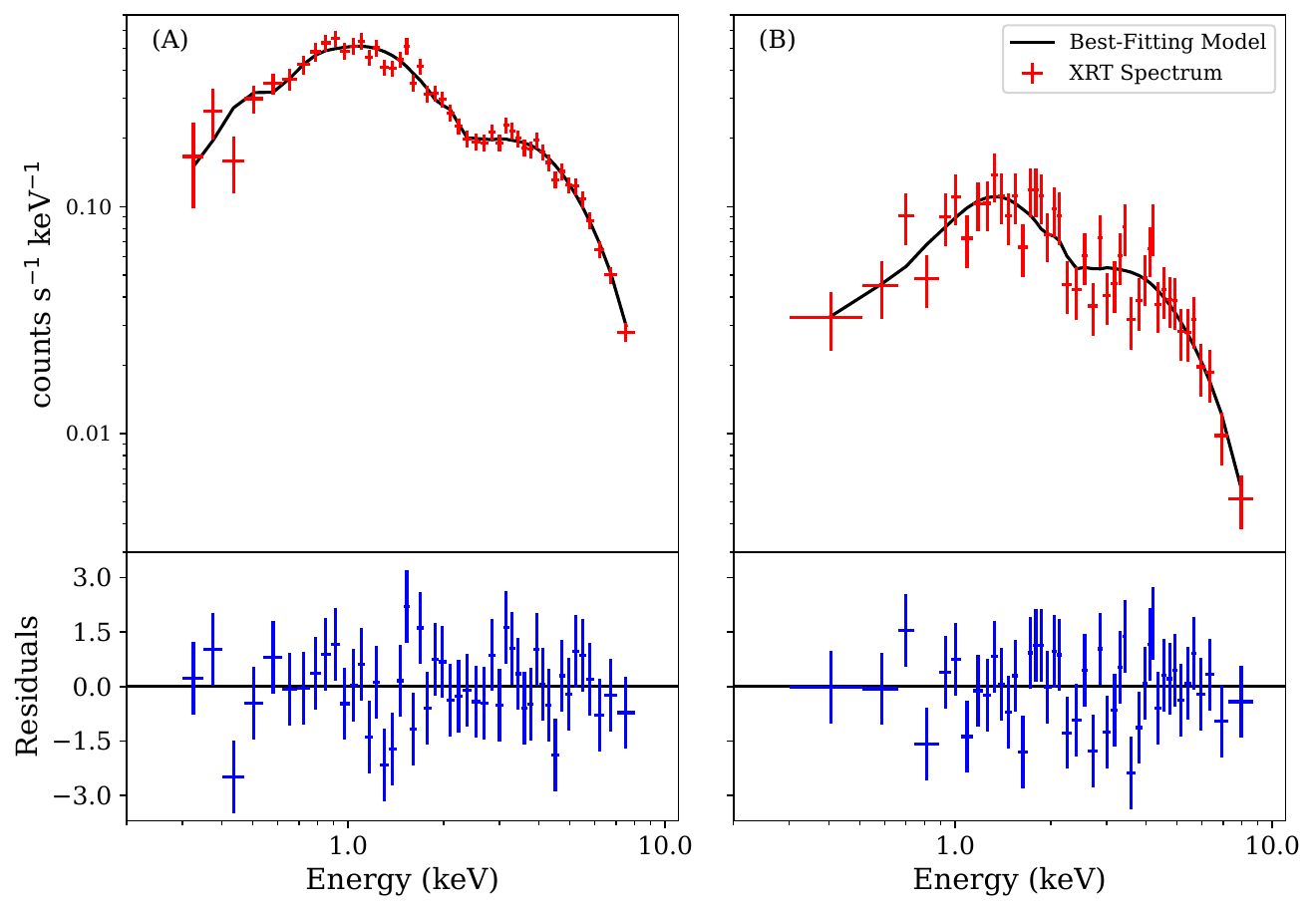}
    \caption{Examples of typical 0.3-10 keV XRT spectra obtained by Swift during the outbursts of \scbe{} plotted alongside the best-fit absorbed power law + blackbody model and the residuals to the fit. \textit{Panel A}: A typical spectrum for the first outburst, obtained by Swift on 2025 June 11. \textit{Panel B}: A typical spectrum for the second outburst, obtained by Swift on 2025 November 17.}
    \label{fig:xrt specs}
\end{figure*}

\begin{figure}
	\includegraphics[width=8cm,angle=-0]{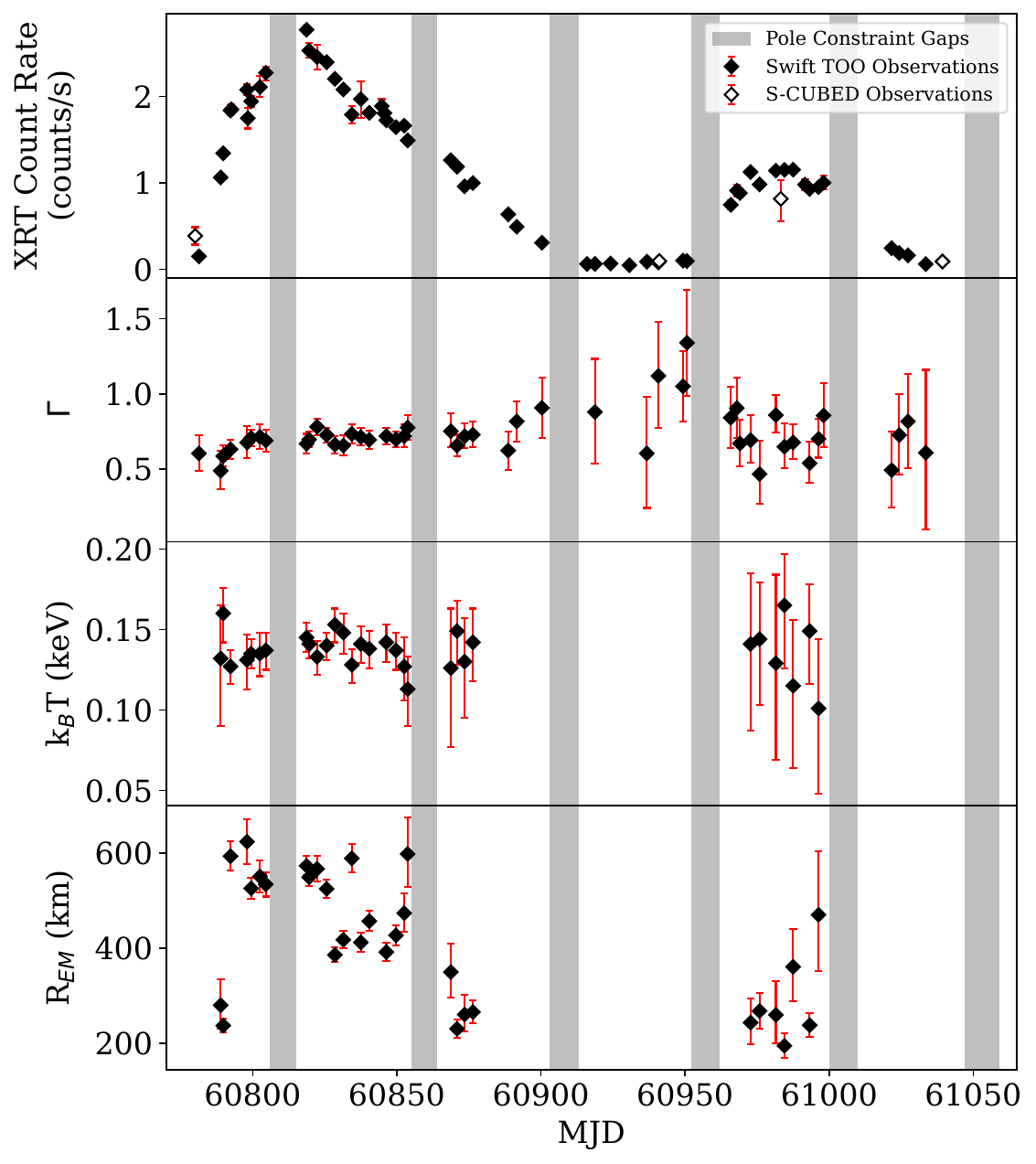}
    \caption{{The variability of the best-fitting XRT spectral model parameters for \scbe{}. Each component of the best-fitting absorbed power law plus thermal blackbody model is shown except for the column density, which was fixed at $1.8 \times 20^{21}$ cm$^{-2}$}. Gray bars represent periods when \scbe{} was not observable by Swift due to Earth limb constraint.}
    \label{fig:xrt spec fits}
\end{figure}

The field containing \scbe{} has regularly been monitored by S-CUBED since the beginning of the survey in 2016, but this source was not detected  in any of the short, 60s XRT exposures until the observation taken on 10 April 2025. The XRT light curve for the duration of S-CUBED is shown in Figure \ref{fig:all_swift}. \scbe{} was first detected during this observation at a count rate of $0.39 \pm 0.1$ counts per second. The source declined in count rate during the first deep target of opportunity (TOO) request that was submitted after detection. However, a second TOO request taken on 23 April 2025 revealed that the source had suddenly increased in brightness and was now detected at an XRT count rate of $1.06 \pm 0.04$ counts per second. 

In response to this sudden increase in count rate, Swift began a regular deep TOO monitoring campaign of \scbe{} that lasted for over 200 days with exposure times ranging from under 1 ks to 5 ks per observation. These observations revealed that the source continued to increase in brightness for 38 days after the initial S-CUBED detection before reaching a peak count rate of $2.77 \, (+0.04, -0.05)$ counts per second. After reaching this peak, the source continued to steadily decline in brightness for over 100 days. At this point, the source entered a plateau phase where the count rate remained at $\sim$0.07 counts per second for over 30 days. Somewhat surprisingly, on 17 October 2025, \scbe{} was observed to have produced a secondary outburst. This secondary outburst peaked at 1.15 counts per second on 8 November 2025. After reaching this peak value, the source began to decay in brightness for a second time. As of the last observation obtained by Swift, this trend continues. The Swift/XRT light curve of both outbursts is shown in the bottom panel of Fig.~\ref{fig:all_swift}.

\begin{table*}
    \centering
    \begin{tabular}{ccccc}
         \hline 
         \hline
         \texttt{xspec} Model & Parameter & Average Outburst 1 Best-Fit Value & Average Outburst 2 Best-Fit Value & Units\\
         \hline
        \texttt{TBabs} & $N_H$\tablefootnote{Column density along the line of sight was fixed at $1.8 \times 10^{21}$ cm$^{-2}$ which is the average value found by \citet{2000Yokogawa} for the 1998 outburst of this source.} & $1.8 \times 10^{21}$ & $1.8 \times 10^{21}$ & cm$^{-2}$ \\
        \texttt{cflux} & $\log_{10}\left(F_{tot, \, 0.3-10 \, \text{keV}}\right)$ & $-9.91\;  (+0.02, -0.02)$ & $-10.28\;  (+0.08, -0.03)$ & erg cm$^{-2}$ s$^{-1}$ \\
        \texttt{powerlaw} & $\Gamma$ & $0.690\; (+0.074, -0.077)$  & $0.700\; (+0.208, -0.220)$ &   -- \\
        \texttt{bbodyrad} & $kT$ & $0.137\; (+0.018, -0.015)$ & $0.135\; (+0.030, -0.025)$ & keV \\
        \texttt{bbodyrad} & $R_{EM}$ & $451\; (+28, -30)$  & $291\; (+39, -44)$ & km \\
        & C-stat (d.o.f.) & 46.5 (41.4) & 28.8 (28) & -- \\ 
        & Null Hypothesis Probability & 0.387 & 0.436 & -- \\
    \end{tabular}
    \caption{Table containing the best-fitting parameters from the absorbed blackbody plus power law model used to describe the spectrum of \scbe{}. Each row describes a free parameter using the model component that it is derived from, its units, and its best-fitting values to the Swift XRT spectra taken throughout the 2025 outburst events. The average best-fitting value is reported for each of the two back-to-back outbursts. These values are consistent with changes to the average flux and the radius of the thermal blackbody emission region during the outburst, but the photon index of the power law and the blackbody temperature do not change between outbursts. \textbf{The default xspec abundances reported in \citet{1989Anders} were assumed for this fitting.}}
    \label{tab:x-ray best fit}
\end{table*}

A 0.3-10 keV spectrum of \scbe{} was obtained for each TOO observation with a duration greater than 1 ks. These observations were obtained with a differing XRT mode depending on the count rate of the source in the previous observation. If the source was detected to be below a count rate of $\sim1.0$ counts per second, then the spectrum was obtained using photon counting (PC) mode. If the source was detected to be above a count rate of $\sim1.0$ counts per second, then the spectrum was obtained using window timing (WT) mode. All spectra were processed using the automated XRT pipeline tools described by \citet{Evans09}. Each one was re-binned using the \texttt{grppha} software package so that each bin had a minimum of 15 counts per bin. However, for deeper observations with large number of counts, the number of counts per bin was increased so that there were between 40 and 50 bins in each spectrum. For the lowest-luminosity observations, a minimum of 4 counts per bin was used. This was the only method that would allow an estimation of the X-ray luminosity of \scbe{} during the periods when the source was at its faintest.

After binning the spectra, the best-fitting spectral parameters were determined for each observation by minimization of the \textit{C}-statistic using the software \texttt{xspec} \citep{Arnaud96}. \textit{C}-statistic minimization is preferred over $\chi^2$ minimization due to its reliability in accurately measuring the parameters of a spectral fit and estimating the uncertainties of these parameters \citep{2017Kaastra}. A typical BeXRB spectrum obtained by XRT is hard and can be fit by an absorbed power law with a photon index of $\Gamma \simeq 1$. However, \scbe{} does not have a typical BeXRB XRT spectrum. Figure \ref{fig:xrt specs} shows a typical spectrum obtained during each outburst. Each of these spectra indicate the presence of a soft X-ray excess over the typical absorbed power law. Because of this soft excess, 33 of the 49 observations are best fit by a multi-component absorbed power law plus blackbody model. The 16 spectra that are not best fit by this multi-component model are best fit by a typical absorbed power law model. It is a lack of detected soft ($< 2$ keV) X-ray photons as the source fades in luminosity that prevents Swift from fitting the absorbed blackbody spectral component of the source during these observations. Once the source re-brightened, the blackbody component was able to be detected again. The fit is also found to be improved when the column density ($N_H$) along the line of sight is fixed at a value of $1.8 \times 10^{21}$ cm $^{-2}$, which is the average value for the column density derived by \citet{2000Yokogawa} using data taken by the Advanced Satellite for Cosmology and Astrophysics (ASCA) during the 1998 outburst of the source. As such, all fits were performed with the column density fixed at this value.

In order to find the radius of emission for the blackbody component in each spectral fit, we followed the method employed by \citet{kennea2021} and \citet{gaudin2024}. After finding the best-fit values for the photon index ($\Gamma$) and blackbody temperature ($kT$), these parameters were then fixed at their best-fit values, and the column density was again fixed at a value of  $1.8 \times 10^{21}$ cm$^{-2}$. The normalization of the blackbody and power law components were allowed to vary using \textit{C}-statistic minimization. The normalization of the \texttt{bbodyrad} component is defined as $n=\frac{R_{EM}^2}{D_{10}^2}$ where $R_{EM}$ is the radius of emission for the blackbody component and $D_{10}$ is the distance to the source in units of 10 kpc. Once a best-fit normalization is found for the source, the radius of emission can be calculated by assuming the standard distance to the SMC of 62 kpc \citep{scowcroft2016}. This makes $D_{10} = 6.2$ which produces a blackbody radius of 400-600 km  during the first outburst and a mean radius of $\sim$290 km during the second outburst (see Figure \ref{fig:xrt spec fits}).

The average best-fitting spectral parameters for our multi-component model during each outburst of \scbe{} are reported in Table \ref{tab:x-ray best fit}. A figure showing the variability of all of the components of our best-fitting model as a function of time is found in Figure \ref{fig:xrt spec fits}. As shown in the figure, the photon index and blackbody temperature parameters are not found to vary significantly as the outburst progresses. Similarly to the 2023 outburst of Swift J0549.7-6812 \citep{2023coe}, there is a slight hint of a softening of the photon index. As these underlying parameters are so steady for the duration of both outbursts, we can conclude that the emission model driving this even does not vary. The only spectral parameter that appears to vary is the radius of emission ($R_{EM}$) of the thermal blackbody component, which increases and decreases with the luminosity of the system by $\sim 300$ km during the first outburst. As expected, during the second and fainter outburst, this blackbody component is found to have decreased in average radius by $\sim$150 km. The average radius of the blackbody component during the second outburst is consistent with the radius observed during the faintest luminosities of the first outburst. 

\subsubsection{UVOT Observations}

\begin{figure}
	\includegraphics[width=8.5cm,angle=-0]{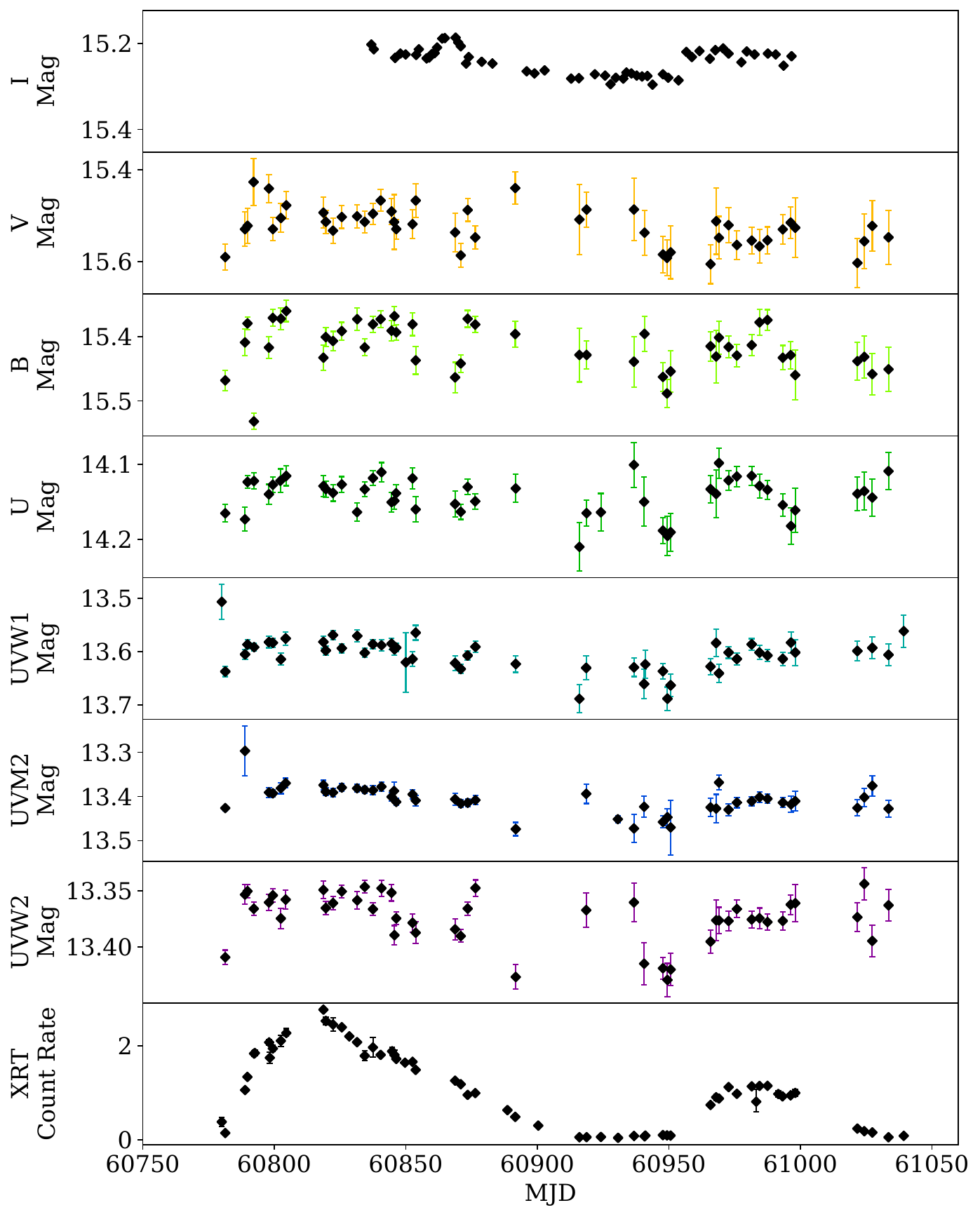}
    \caption{{Multiwavelength light curve spanning the entire duration of the 2025 outburst for \scbe{}. OGLE and Swift UVOT band light curves are plotted alongside the 0.3-10 keV XRT count rate of the source, showing that the optical/UV brightness of the source appears to change with the X-ray count rate.}}
    \label{fig:all_swift_outburst}
\end{figure}

\begin{figure}
	\includegraphics[width=8cm,angle=-0]{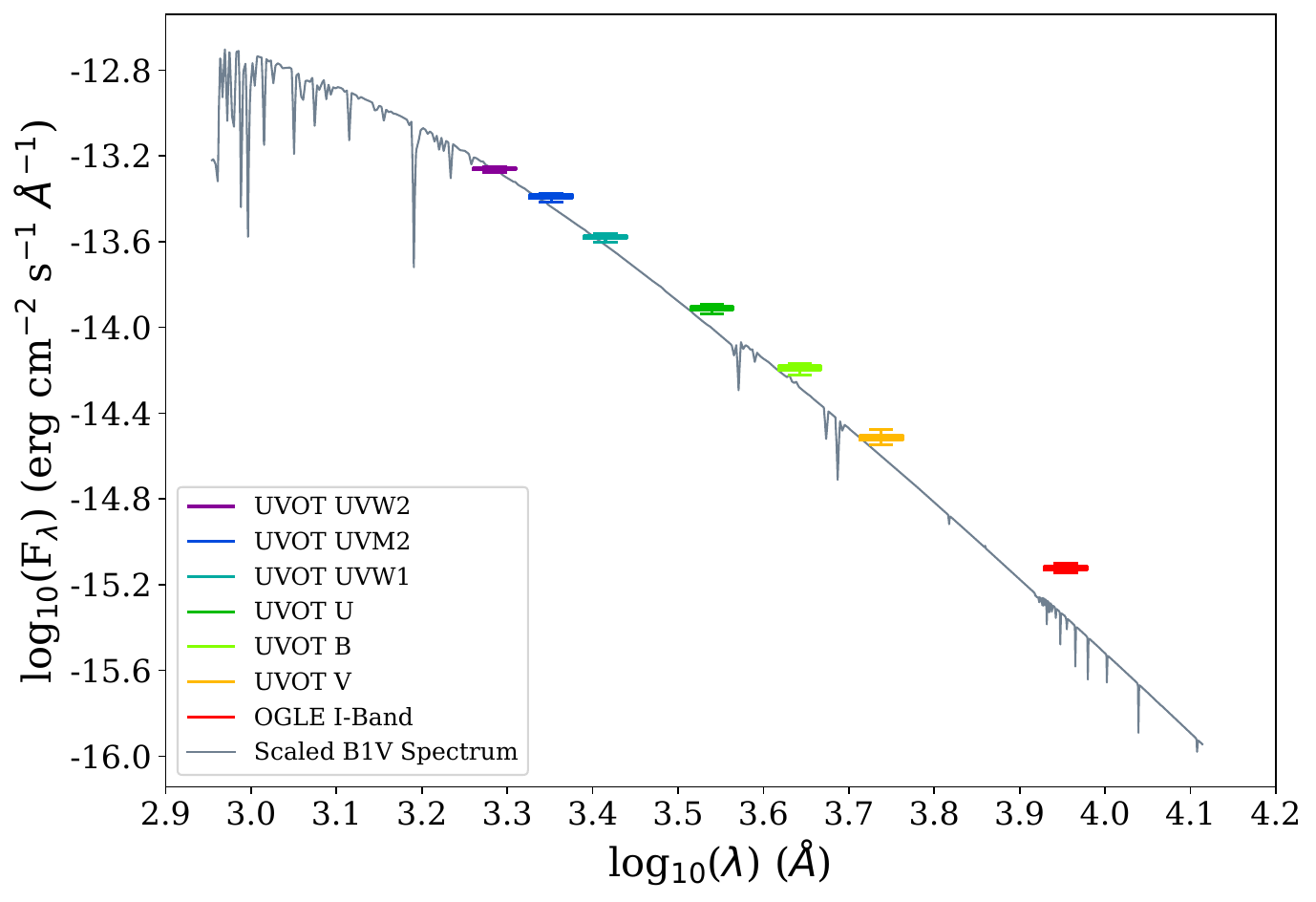}
    \caption{{Average flux for \scbe{} during the 2025 outburst in each of the Swift UVOT photometric bands as well as in the OGLE \textit{I}-band. The variability is plotted for each band as a box-and-whisker plot. Also plotted is a PHOENIX model spectrum of a standard B1V star scaled to the median flux of the \textit{uvw2}-band for comparison. }}
    \label{fig:uvot_sed}
\end{figure}

\scbe{} has been detected by UVOT consistently since the start of the S-CUBED survey. A {uvw1}-band light curve with approximately weekly coverage was generated for this source via aperture photometry using the FTOOLS \citep{1999Blackburn} module \texttt{uvotsource}. In order to properly estimate the magnitude of the source in each observation, the circular source aperture radius that is used by \texttt{uvotsource} was set to 5$''$, and a nearby ($\lesssim 30''$) circular background region was selected to have a radius of 8$''$ and to contain no stars. Using this method, the light curve presented in the middle panel of \ref{fig:all_swift} is shown spanning the nearly 10 years of S-CUBED.

This light curve shows relatively unusual behavior for a BeXRB in the SMC. Instead of demonstrating variability that is approximately correlated with the behavior that is observed in the OGLE I-band (top panel of \ref{fig:all_swift}), there is little evidence for variability of the source at all. The mean \textit{uvw1}-band value of the pre-outburst S-CUBED observations for SXP 31.0 is found to be 13.6 magnitude, and all variability is consistent with the measurement noise associated with short, 60s S-CUBED exposures. 

The lack of observed \textit{uvw1}-band variability is a surprising result that diverges from the typical behavior observed in other active SMC BeXRBs. In many cases \citep{Kennea2020, 2024Gaudin}, S-CUBED has observed correlated variability between the \textit{uvw1}-band and the \textit{I}-band light curves of a system on super-orbital or multi-year timescales. In this case, the \textit{uvw1}-band data does get brighter or dimmer as the I-band magnitude of the system changes with time. Instead it remains constant, suggesting that there is no variability in the circumstellar disk contribution to the UV flux of the system. 

During the X-ray outburst, deeper exposures reveal that there is a small amount of variability that can be observed in the \textit{uvw1}-band. \scbe{} is found to vary in brightness by up to 0.1 mag during the >200 day span of the outburst. Additionally, this variability is found to be correlated with the variability of the \textit{I}-band data and the XRT count rate. This correlation during large outbursts has been observed before in systems such as SMC X-3 \citep{Townsend2017}. 

The deeper TOO observations of the source during outburst were taken with UVOT in 0x30ed mode, which is a uv-weighted filter that observes the target with each of Swift's 6 UV/optical filters. Individual snapshots for the sky image and exposure map of a given observation and filter were summed together using the \texttt{uvotimsum} command after removing snapshots for which small-scale sensitivity issues were observed. The method described above was then used to perform 6-filter aperture photometry on the target. Multiwavelength aperture photometry resulted in the light curve that is shown in Fig.~\ref{fig:all_swift_outburst}.

In this multiwavelength light curve, we see evidence of coherent variability in all bands observed by Swift, but the evidence for an associated optical flare that accompanies the large X-ray outburst is weaker due to the greater uncertainty associated with these observations. No UVOT filter sees the magnitude of the source change in brightness by more than 0.2 magnitudes, and the change in magnitude appears to decrease as you progress to shorter wavelengths. Again, this is a somewhat surprising result. There is a growing body of evidence suggesting that an IR-UV flares should accompany large X-ray outbursts due to the X-ray heating of the circumstellar disk by the NS companion \citep{2024Alfonso-Garzon, 2025Vasilopoulos}. However, NS heating is challenged by the small variability that is observed in the extensive UVOT monitoring that has been performed for \scbe{}. It is possible that the NS is not at a favourable angle to produce this disk heating effect or that it never makes a close enough approach to the disk for the hotter material in the innermost regions of the disk to be strongly effected. More investigation is needed to better understand this effect and the small magnitude of the  associated flare in this instance.  

The lack of variability can be further underscored by creating a time-averaged spectral energy distribution (SED) for our source using the Swift UVOT data and OGLE \textit{I}-band data. In addition to producing magnitudes for each observation and filter, \texttt{uvotsource} also produces simultaneous flux measurements. These values were de-reddened for each filter using the SMC reddening maps from \citet{skowron2021} and the methods outlined in Section 3.2 of  \citet{2025Gaudin}. Once reddening was accounted for, a box-and-whisker plot was made for the data from each filter. All box-and-whisker plots were plotted at the respective central wavelengths of the Swift UVOT filters in Fig.~\ref{fig:uvot_sed}. Plotted alongside them is a Phoenix model \citep{2013Husser} of a B1V star that is scaled to the median value for the flux of the \textit{uvw2}-band filter. This plot suggests that the variability of the source appears to decrease at shorter wavelengths, suggesting that the majority of the changes in the size and structure of the circumstellar disk happen at larger radii where the disk is cooler. 

\subsection{Optical photometry}
\subsubsection{OGLE}
The OGLE project \citep{Udalski2015} undertakes to provide long term I-band photometry with an average cadence of 1-3 days. The optical counterpart to \scbe{} ~was observed continuously for over 2 decades in the I-band with only a gap of $\sim 2.5$ year due to Covid-19 restrictions. After that gap, the source was observed with a higher cadence for the following seasons. 

The source falls upon 2 separate chips in the OGLE survey so it has several identities in the OGLE catalogues - see Table~\ref{tab:OGLE_ID}

\begin{table}
	\centering
	\caption{OGLE source identifications}
	
    \setlength\tabcolsep{2pt}
	\begin{tabular}{ccc} 
		\hline\hline
		Band & OGLE III name(s) & OGLE IV name(s)   \\
		\hline

I & smc726.08.69 \& smc733.32.102 & smc726.08.69 \& smc733.32.102 \\
V & smc116.6.v.17 & smc726.08.v.31 \& smc733.32.v.32 \\

		\hline
	\end{tabular}
\label{tab:OGLE_ID}	
\end{table}

\begin{figure}
	\includegraphics[width=8cm,angle=-0]{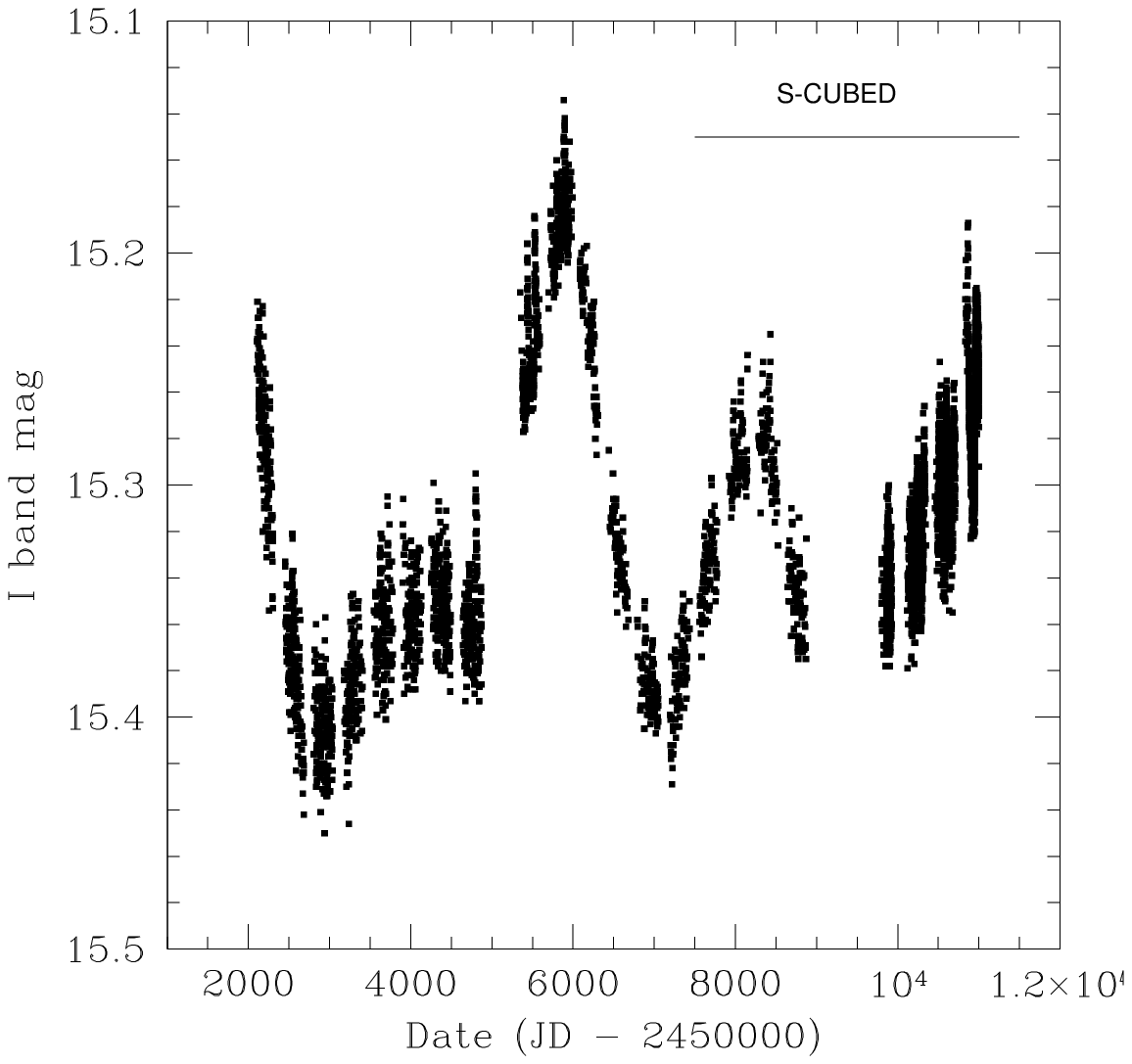}
    \caption{{24 years of OGLE III and IV observations of the optical counterpart to \scbe{}.} The gap in the OGLE IV coverage for the period JD 245889 - 2459800 corresponds to the time of Covid-19 telescope closure. The duration of the S-CUBED project is shown. }
    \label{fig:ogle}
\end{figure}

The 24 years worth of I band data are shown in their entirety in Fig.~\ref{fig:ogle}. From this figure it is clear that the optical signature of the system is extremely variable on timescales of several years. This is classic behaviour for Be stars (e.g. \citealt{Rajoelimanana2011}) with the variability being indicative of changes in the mass outflow from the star feeding a circumstellar disc of varying size.

The date of the first known X-ray outburst 1998 Oct - 1999 Feb falls before the start of the OGLE III coverage which began in 2001 July. Subsequently, the brightest observed peak in the combined OGLE III \& IV data (I$\sim$15.15) is seen to occur around JD 245600 - 2012 March. If an X-ray outburst was triggered at this time it has not been reported. The second brightest OGLE I band peak (I$\sim$15.20) presumably triggered the current X-ray outburst that is the subject of this paper.

\subsubsection{Period analysis}
\begin{figure}
	\includegraphics[width=8cm,angle=-0]{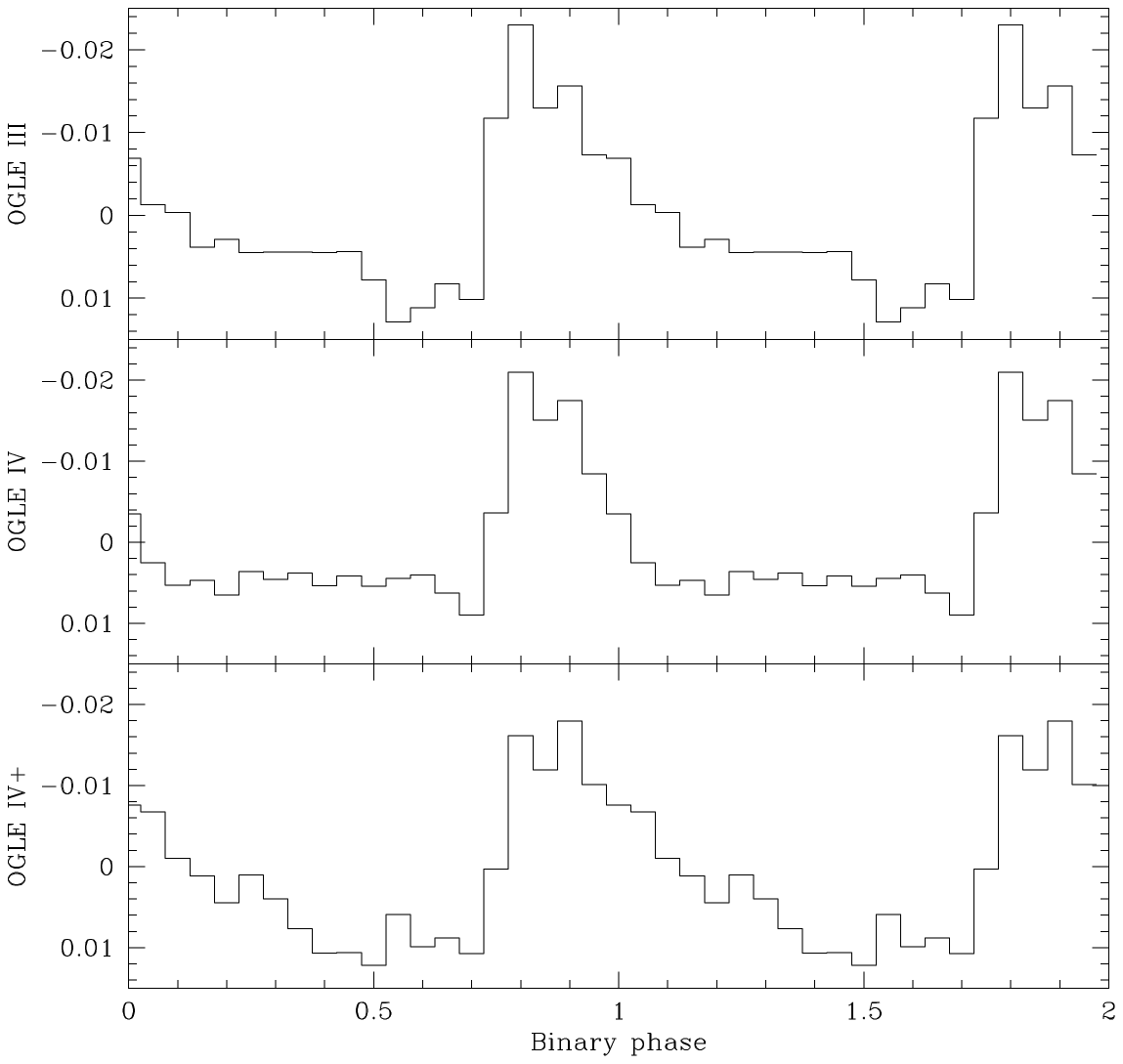}
    \caption{The detrended OGLE data divided into three phases and all folded at the same period of 90.53~days and with the same zero point of JD 2452104.0. From top panel to bottom they show OGLE III, OGLE IV (pre-Covid-19) and OGLE IV+ (post Covid-19).The lowest panel includes the time of the X-ray outburst.}
    \label{fig:folds_all}
\end{figure}

The detrended OGLE III \& IV data were searched for possible periodicities in the range 2 - 200d using a Generalized Lomb-Scargle technique \citep{zk2009}. In all segments a significant period was detected with a value of 90.5 $\pm$ 0.2 days. \cite{bird2012} found a period of 90.53 $\pm$ 0.07d from analysing the all OGLE II and OGLE III data sets.
So the OGLE data from before and after the Covid-19 gap were separately folded at the period of 90.53d and are shown in Fig.~\ref{fig:folds_all}. Though the fundamental period is unchanged the folded light curve sometimes exhibits a much broader structure than on other occasions. See the discussion (Section~\ref{discussion}) for possible interpretation of this change.

Based upon the OGLE III data the ephemeris for the brightest point in the 90.53d cycle is given by: 

\begin{equation}
T_{\mathrm{peak}} = 2452176.4 + N(90.53)~\textrm{JD}
\label{eq:1}
\end{equation}

It is probable that this optical peak position is close to the phase of the periastron passage of the neutron star, and is indicative of the surface area of the circumstellar disc being marginally increased at this orbital phase.

\subsection{SALT spectroscopic observations}

\scbe{} was observed using the Southern African Large Telescope (SALT) with the Robert Stobie Spectrograph (RSS; \citealt{Burgh03}). Two grating settings were employed: PG0900 with a resolution of approximately 6~\AA~ (wavelength range: 5040 -- 8060~\AA) and PG2300 with a resolution of approximately 1.7~\AA~ (wavelength range: 6090 -- 6920~\AA). The PG0900 and PG2300 observations were obtained using exposure times of 1200~sec and 2000~sec, respectively. 
An observation using the High Resolution Spectrograph (HRS) was conducted in low-resolution mode (R $\sim$ 14,000). The observation consisted of a single exposure lasting 2400 seconds and covered a wavelength range of approximately 5500–8800~\AA, with a resolution of about 0.4~\AA. A summary of the observations is presented in Table~\ref{tab:salt_tab}.

The first observation performed with the PG0900 grating revealed the H$\alpha$ line in emission, with a measured equivalent width (EW) of -32.44$\pm$0.69~\AA ~and likely originates from a combination of contributions from the Be disc (broad wings) and circumstellar material (central narrow emission). Previous published values for the EW were -27.0$\pm$0.3~\AA ~on 9 January 1999 from SAAO \citep{coe2003} and -21~\AA ~later the same month from ESO \citep{covino2001}. Those latter authors also report further observations on 15 Sept 1999 and a much higher value of -40~\AA. 
\begin{figure}
	\includegraphics[width=8cm,angle=-0]{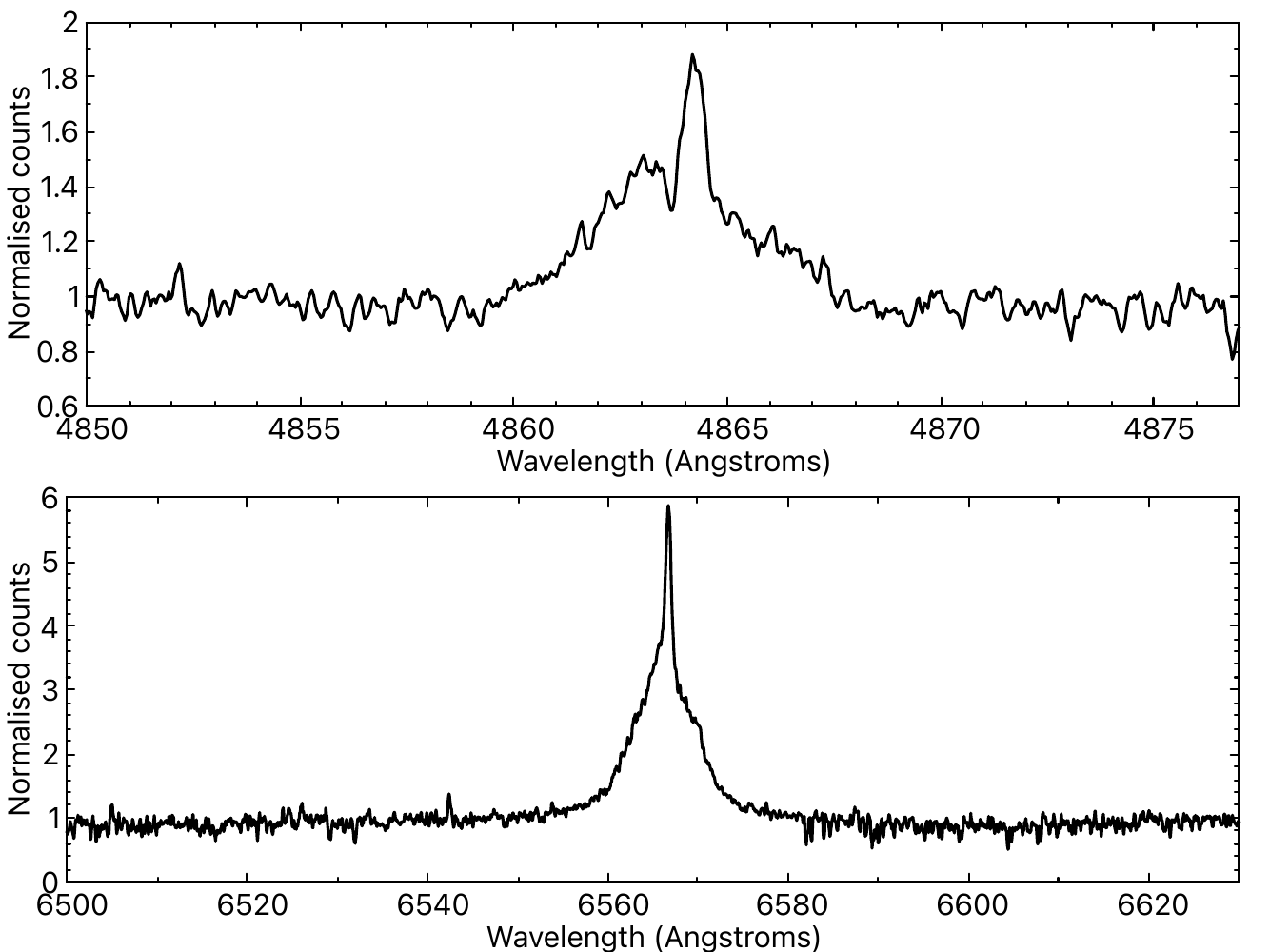}
    \caption{The H$\beta$ (top) and H$\alpha$ (bottom) line profiles for \scbe{} from the SALT HRS observation of 20 November 2025.}
    \label{fig:SALT_evo}
\end{figure}
Subsequent SALT spectra were obtained using the higher-resolution PG2300 grating and one using the HRS. These higher-resolution observations reveal a multi-component emission line, which displays a narrow emission component superimposed on a broader one. The narrow emission component is likely derived from the extended emission surrounding the target, while the broader component likely originates from the Be disc. 

We performed a multiple-peak fit to the higher-resolution profiles, employing a Voigt profile for the broad component and a Gaussian fit for the narrow component. This analysis yielded equivalent widths of approximately $-5$~\AA~ for the narrow component, resulting in total EWs slightly less than the measured value of -32.44$\pm$0.69~\AA~ from the first observation taken on July 13, 2025 (MJD60870.149120).

\begin{table}
	\centering
	\caption{A log of all the SALT RSS long slit observations. The H$\alpha$ EW measurements of \scbe{} are presented in this table.}
	\label{tab:salt_tab}
    \setlength\tabcolsep{2pt}
	\begin{tabular}{cccc} 
		\hline\hline
		Date & MJD (orbital phase) & EW (\AA) & Grating  \\
		\hline

13 July 2025 & 60870.149120 (0.037) &	-32.4$\pm$0.7 & PG0900 \\
06 August 2025 & 60894.096956 (0.302) &	-27.4$\pm$0.6 & PG2300 \\
10 August 2025 & 60898.117500 (0.346)  & -28.3$\pm$0.4 & PG2300 \\
14 August 2025 & 60902.120567 (0.390)  & -28.7$\pm$0.9 & PG2300 \\
17 August 2025 & 60905.076562 (0.423)  & -29.6$\pm$1.4 & PG2300 \\
20 August 2025 & 60908.023692 (0.456)  & -27.1$\pm$1.4 & PG2300 \\
03 November 2025 & 60982.82509 (0.282)  & -30.9$\pm$0.2 & PG2300 \\
05 November 2025 & 60984.81180 (0.304)  & -28.8$\pm$0.2 & PG2300 \\
11 November 2025 & 60990.90550 (0.371)  & -32.1$\pm$0.7 & PG2300 \\
20 November 2025 & 60997.85360 (0.371)  & -31.6$\pm$1.5 & HRS \\
		\hline
	\end{tabular}
\label{tab:SALT}	
\end{table}

\begin{figure}
	\includegraphics[width=8cm,angle=-0]{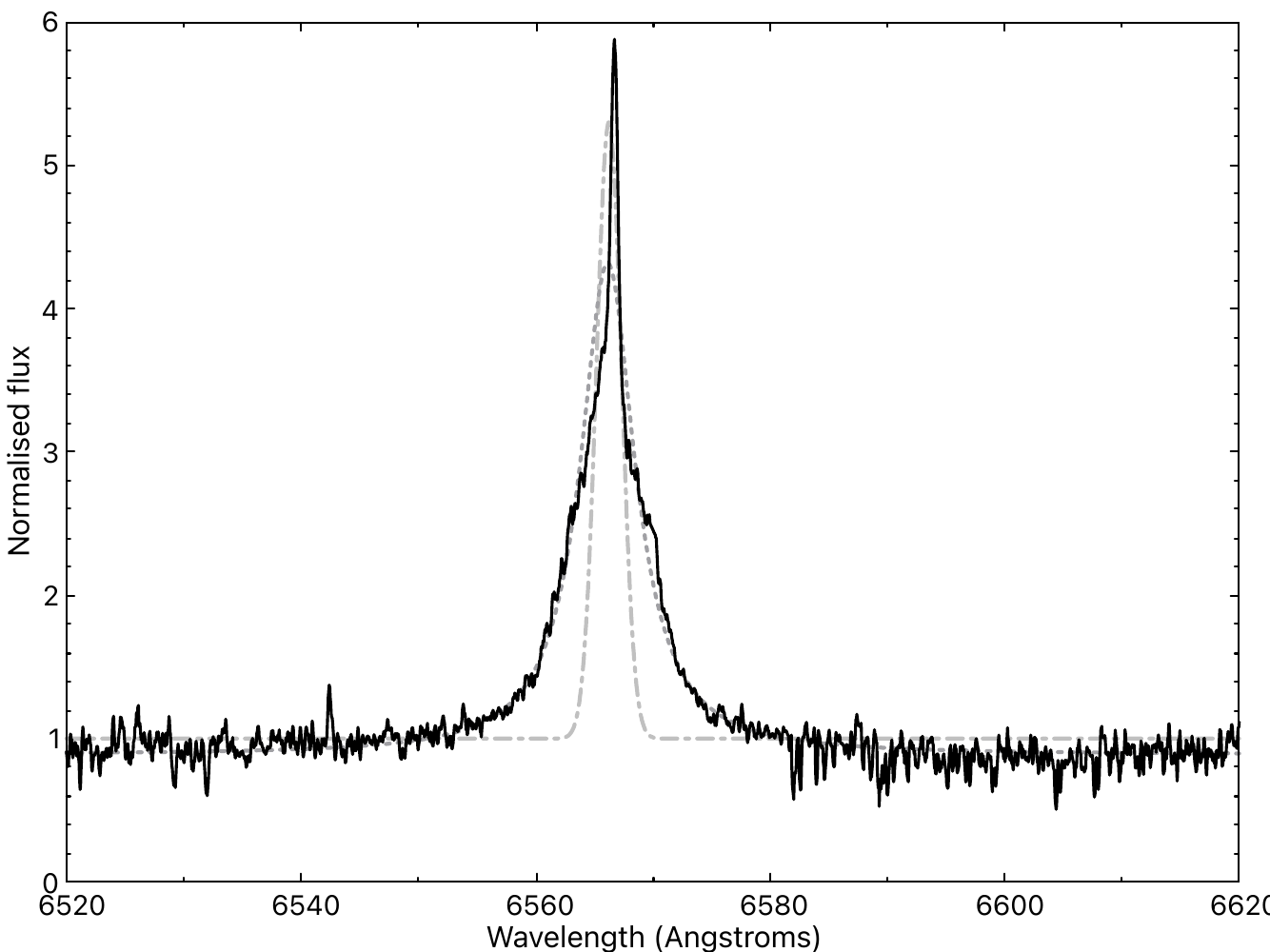}
    \caption{The H$\alpha$ emission line obtained with HRS (solid line) from 20 November 2025 showing the combination of the Gaussian (dashed-dotted line) and Voigt (dashed line) fits.}
    \label{fig:SALT}
\end{figure}

The H$\alpha$ spectrum obtained with HRS and the 2-model fit is shown in Fig.~\ref{fig:SALT}. The possible difficulties in obtaining an accurate measurement for the star's H$\alpha$ EW in the presence of a surrounding H$\alpha$ nebulosity is discussed below.

\subsection{SALT Imaging observations}

\begin{figure}
	\includegraphics[width=8cm,angle=-0]{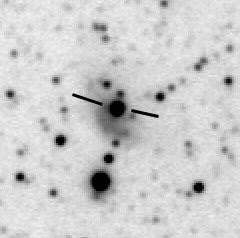}
    \caption{H$\alpha$ image from 6 August 2025. The image size shown is approximately 70 x 70 arcsec. The position of \scbe{} ~is indicated.}
    \label{fig:SALTICAM}
\end{figure}

It has been reported that there is extended emission clearly visible in the H$\alpha$ \citep{coe2003}. So, in order to hopefully better understand the reliability of the H$\alpha$ flux measurements, an H$\alpha$ image was obtained using SALTICAM. A dither pattern of 12 exposures was used, each of 100s.
The resulting image obtained by combining all 12 exposures is shown in Fig.~\ref{fig:SALTICAM}.

\begin{figure}
	\includegraphics[width=8cm,angle=-0]{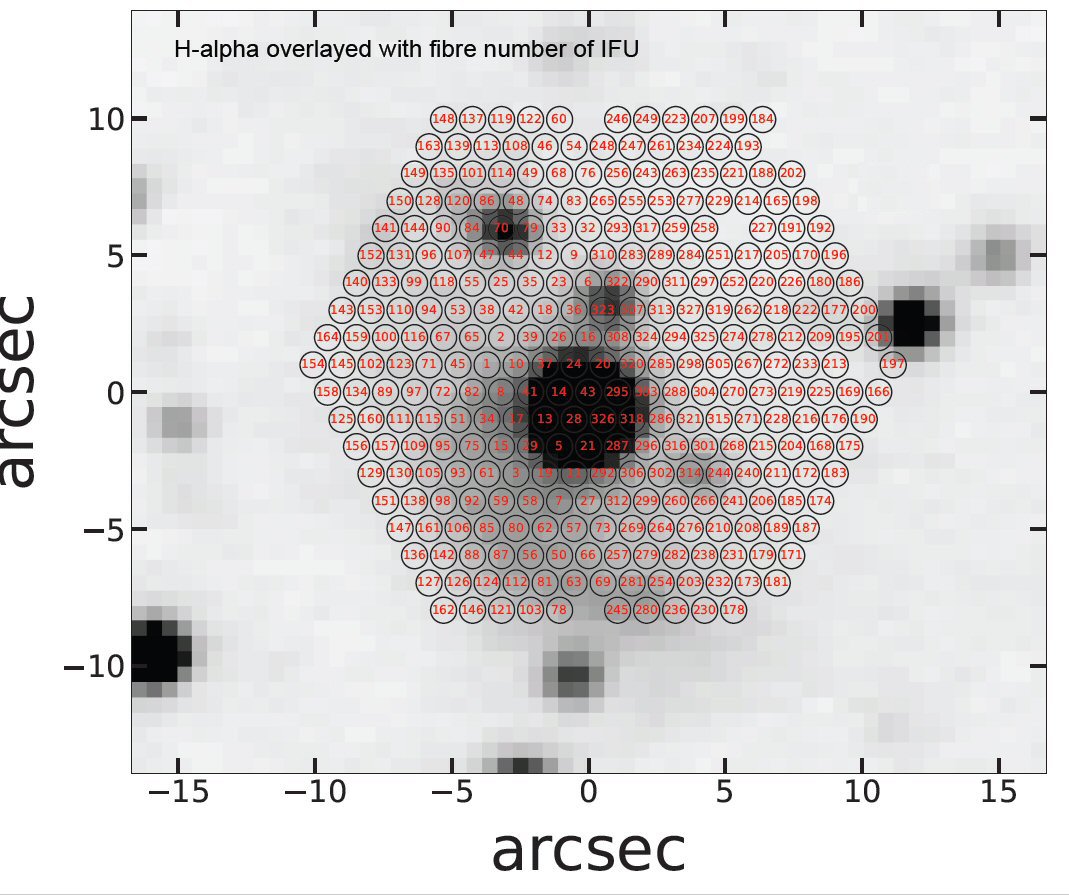}
    \caption{The layout of the fibres from the SMI200 observation supermposed on the H$\alpha$ image of \scbe{} - see Fig.~\ref{fig:SALTICAM}. The fibre array's size is 22.5 x 17.6 arcsec. }
    \label{fig:IFU}
\end{figure}

In order to investigate the nature of the halo surrounding \scbe{} ~observations were carried out on 20 August 2025 using SMI200 - the Slit Mask Integral Flux Unit on SALT \citep{Chattopadhyay}. This instrument has a total of 327 fibres which cover a sky area of 22.5 x 17.6 arc seconds. There are 24 sky fibres and the rest are arranged in an extended hexagonal pattern. These were located on the sky such that individual parts of the H$\alpha$ emitting region surrounding \scbe{} could be spectroscopically investigated - see Fig.~\ref{fig:IFU}.

The seeing conditions were $\sim$1.3 arcsec and fibre cores are 0.88 arcsec in diameter, but the centre-to-centre separation between fibres is 1.06 arcsec. So a stellar point source will be spread over a few fibres. The PG0900 grating was used with a grating angle of 15.125$^{\circ}$ and the exposure time was 2400 seconds for each of the two frames obtained. The resulting spectral coverage was 4200 -- 7200 \AA.


The conclusions from all the imaging observations are presented in the discussion (Section 3.6).


\section{Discussion}\label{discussion}

\subsection{The X-ray outburst trigger} \label{subsec:x-ray trigger}

\begin{figure}
	\includegraphics[width=9cm,angle=-0]{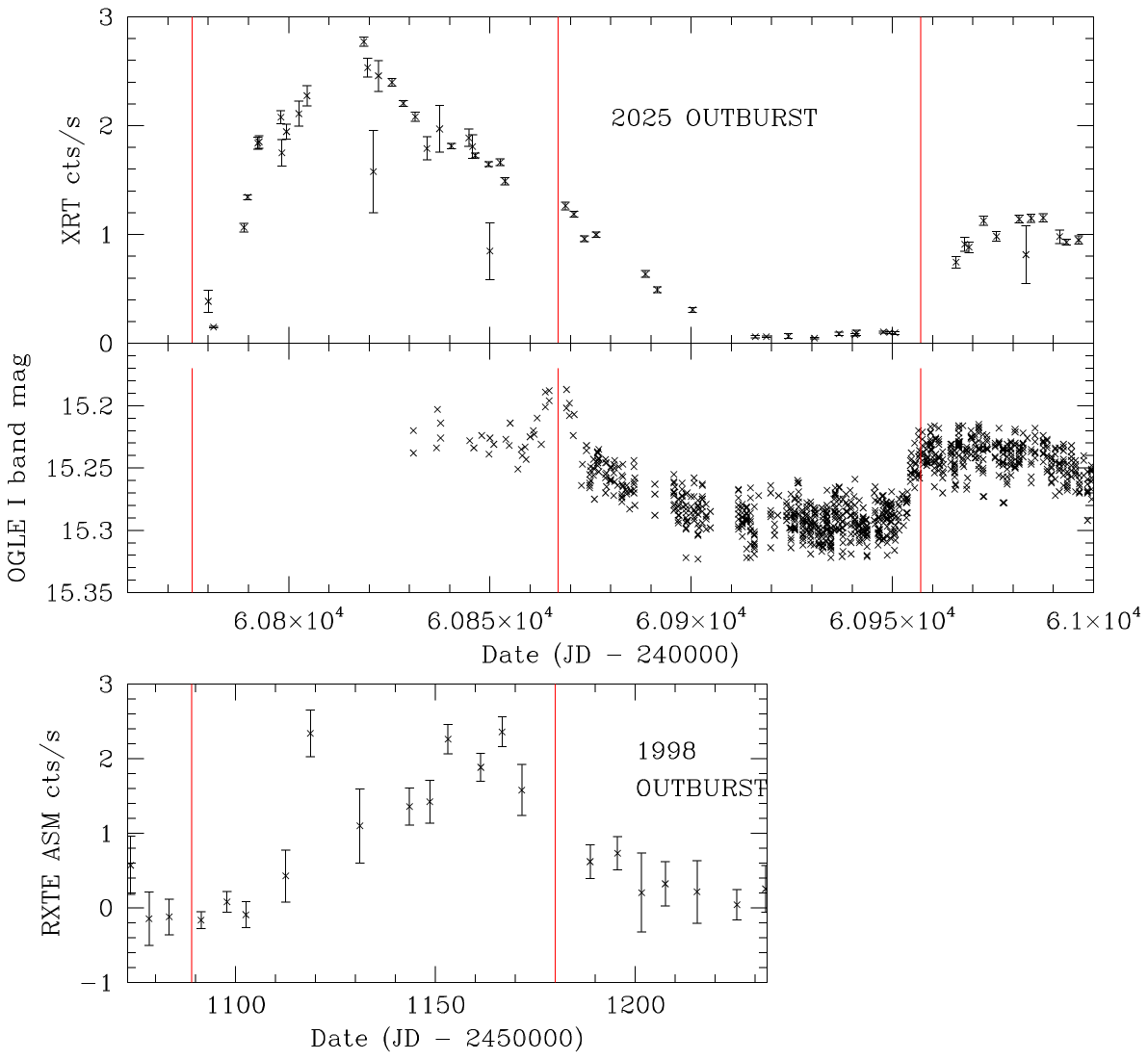}
    \caption{Comparison of the X-ray outburst profiles from the current outburst (top panel) and the 1998-9 one (bottom panel). Also shown are the I band measurements from the current outburst (middle panel).  The vertical red lines show the predicted times of OGLE peaks based upon Equation 1.}
    \label{fig:xoa}
\end{figure}

Fig.~\ref{fig:xoa} shows the SXP 31.0 outburst profiles in the X-ray (0.3-10 keV) and OGLE I band. The red lines indicate the predicted I band peak time based upon fits to all the previous $\sim$20 years of OGLE data - see Equation 1. The 2-10 keV X-ray data from the previous event (1998-9) come from the archive produced by the All Sky Monitor instrument on the Rossi X-ray Timing Explorer (RXTE/ASM) \citep{levine1996}. 

It is very possible that the I band peak time corresponds to the periastron passage of the NS.
This would be consistent with the circumstellar disc being maximally disrupted at this time of closest neutron star approach, and thereby temporarily increasing its surface area. However, the X-rays do not seem to indicate any obvious reaction to such a periastron moment, unless the start of the two X-ray outbursts was triggered around that point in time.

More generally, the overall profiles of the two outbursts are very similar. Both last about $\sim$100 days and appear to peak at similar luminosities. The quoted peak X-ray luminosity of $1\times10^{38}$ erg/s was reported for the 1998 outburst \citep{chak1998a}. See next section for more detailed determination of the X-ray luminosity in the 2025 outburst.

\subsection{The X-ray luminosity evolution of the outburst} \label{subsec:x-ray lum}

\begin{figure}
	\includegraphics[width=8cm,angle=-0]{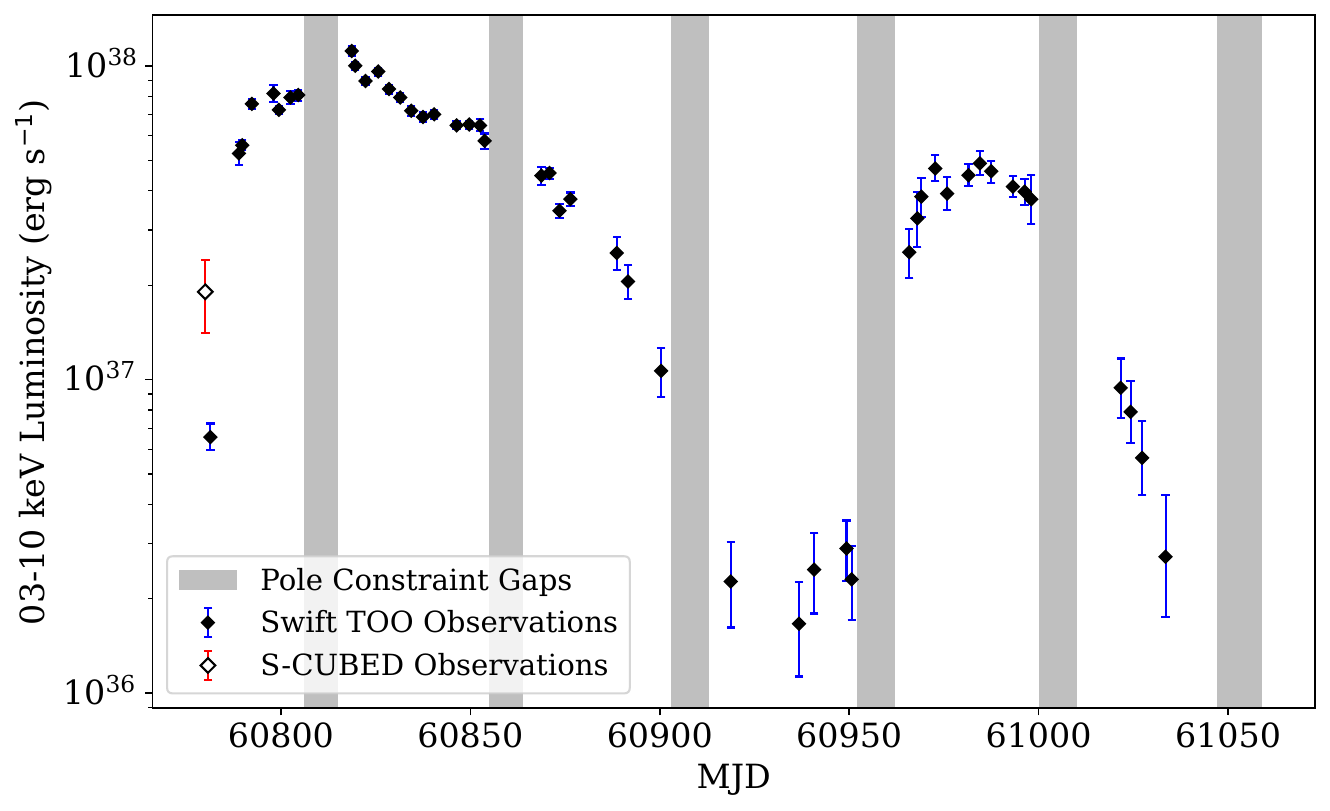}
    \caption{{XRT luminosity lightcurve for \scbe{}. The 60 second S-CUBED observations are shown with red vertical error bars. Deeper target of opportunity Swift observations are shown with blue error bars. Gray boxes indicate the periods of time when the source was unobservable due to Swift Earth Limb constraints.}}
    \label{fig:xrt}
\end{figure}

Monitoring the X-ray luminosity of SXP 31.0 during this outburst is a non-trivial exercise due to the multi-component nature of the spectrum. Typically, the X-ray luminosity of a BeXRB outburst can be calculated by assuming an absorbed power-law spectrum and deriving a counts-to-flux ratio for the outburst from this constant underlying spectrum that can be applied to each observation. However, the thermal component discussed in Section \ref{subsubsec:swift_xrt} complicates this method. According to Fig.~\ref{fig:xrt spec fits}, the thermal component itself appears to be constant throughout the outburst, but this component is faint and can only be detected when there are a large number of counts available in the soft energy bands (0.3-2.0 keV) of XRT. Therefore, the luminosity of the source cannot be calculated by assuming the same model and deriving a counts-to-flux ratio. Instead, it must be determined by fitting each individual spectrum to multiple models (thermal+power-law and just power-law), determining both which model fits the data best, and arriving at the best-fitting model parameters for the spectrum. In some cases, either the exposure time of the observation or the count rate of the source is too low for spectral fitting to be possible. In these cases, no luminosity is reported for the source. For all other observations, the 0.3-10 keV X-ray luminosity evolution of the outburst is shown in Fig.~\ref{fig:xrt}.

According to this figure, the luminosity of the source increases rapidly over the first 10 days of the outburst. During these first few days, there is almost an order of magnitude increase in luminosity during this rapid brightening phase. SXP 31.0 then spends the next month slowly increasing in luminosity until it reaches its peak value of $L_{X} = 1.13\; (\pm \; 0.04) \times 10^{38}$ erg s$^{-1}$ on 23 May 2025. This peak luminosity corresponds to a near-Eddington luminosity of $L_{X} \sim 0.59 L_{Edd}$. After reaching a near-Eddington luminosity, the source decays very slowly, remaining above $L_{X} \approx 2 \times 10^{37}$ erg s$^{-1}$ for almost 3 months. It is not until 13 Aug 2025 that the source begins to rapidly decay in luminosity, dropping in luminosity by over a factor of 10 in under a month. 

This rapid decay would normally indicate the beginning of the propeller effect regime where the accretion rate is no longer high enough to overcome the rapidly rotating magnetic field surrounding the NS. However, instead of a rapid return to quiescence that normally accompanies the onset of the propeller regime, the system experienced a plateau phase lasting for longer than a month from mid-August to mid-September of 2025. During this period, the luminosity remains constant at an XRT luminosity of $L_{X} \approx 2 \times 10^{36}$ erg s$^{-1}$.

After experiencing a month-long plateau phase from August to September of 2025, it was expected that the source would fade in X-ray luminosity below the detection threshold of Swift. Instead, when the source emerged from a 10-day Swift observational constraint gap, the source had increased again in X-ray luminosity by an order of magnitude to $L_X = 2.8 \times 10^{37}$ erg s$^{-1}$. Swift then observed the source to increase rapidly in X-ray luminosity to $L_X \approx 4.5 \times 10^{37}$ erg s$^{-1}$ before beginning to decay with a similar shallow profile as was observed during the initial outburst. This second outburst is coincident with an optical re-brightening seen in the OGLE data consistent with Eqn.~\ref{eq:1} - see Fig.~\ref{fig:xoa}.

\subsection{An unusual Type II outburst pair}

As noted in Sections \ref{subsec:x-ray trigger} and \ref{subsec:x-ray lum}, the two X-ray outbursts of this source have remarkably similar profiles and peak luminosities. While it is not known yet exactly how long this second outburst will last before it begins to decay, it has already achieved a duration of 45 days which is half of the proposed orbital period of the system. Given the commonly-used classification criteria of X-ray outbursts in BeXRB systems (e.g. \citealt{2001Okazaki}), the parameters of this second outburst necessitate its classification as a second Type II outburst. 

Type II outburst ``pairs" have been observed before in BeXRB systems. However, the properties of these paired systems are much different than what is observed in SXP 31.0 The most prominent example of a system that produces paired outbursts is the galactic BeXRB 4U 0115+63 \citep{2018Reig}, which produces frequent outburst pairs. Another recent example from the SMC is SXP 8.80 \citep{2025bGaudin} which was found by S-CUBED to produce Type II outbursts in 2023 and 2025. According to theoretical models of paired outburst systems \citep{2019Martin, 2021Franchini}, the paired outburst events are expected to be separated by a gap of $\sim$1.5 - 3 years of quiescence and experience a decrease in maximum X-ray luminosity from the first to the second event. This type of event is thought to be triggered by the Von Zeipel–Lidov–Kozai (ZLK) mechanism in which interactions between the warped circumstellar decretion disk of the Be star and the accretion disk surrounding the NS produce a highly eccentric accretion disk that can produce a re-brightening of the system after several binary orbits. 

SXP 31.0 does not follow the observational parameters of other BeXRBs that produce outburst pairs. While we do see a second outburst with a lower X-ray luminosity than in the original one, the time separation between these outbursts is only 45 days, which corresponds to approximately half of one orbital period. To the best of our knowledge, this is the shortest time separation ever observed between Type II outbursts in a BeXRB system, making the paired outburst an extremely unusual event.

The duration of 45 days is not a long enough time for binary interactions to induce the eccentric accretion disk that would be required to drive a re-brightening event. Instead, the coincidence of this re-brightening with a proposed periastron passage of the system suggests that the material in the circumstellar decretion disk is responsible for both events. If the material in this disk was not significantly disrupted by the original Type II outburst, then it is possible that it was still large enough to interact with the NS on its next periastron passage. This led to even more material being stripped from the disk to fuel accretion onto the NS, producing a second outburst in quick succession. 

\subsection{Be circumstellar disc}

It was demonstrated in \cite{coe2015} that a relationship generally exists between the size of the circumstellar disc and that of the neutron star orbit. The latter possibly constraining the growth size of the disc.

A direct test of this proposed physical link between
the size of the neutron star orbit, $a$, and the size of
the circumstellar disk, $R_{cs}$, would be to determine both of
these parameters for \scbe{}. To determine the minimum size of the circumstellar disc the smallest H$\alpha$EW value presented in Table~\ref{tab:salt_tab} was used (-27 \AA) and inserted into the relationship from \cite{Hanuschik1989}:

\begin{equation}
log(\sqrt(\frac{R_{OB}}{R_{{cs}}})= [-0.32\times log(-EW)]-0.2
\end{equation}

In this expression $R_{OB}$ is the radius of the Be star (a spectral type of a B1V is assumed and a value of $5 \times 10^{9}$m), and $EW$ is the average H$\alpha$ EW values in \AA.

The size of the semi-major axis, $a$, of the neutron star's orbit may be determined from Kepler's Third Law:

\begin{equation}
\label{eqn:kepler}
a=[\frac{P_{orb}^{2}G(M_{ns}+M_{OB})}{4\pi^{2}}]^{1/3}
\end{equation}

where $M_{OB}$ is the mass of the specific Be star, $M_{ns}$ is the mass of the neutron star (assumed here to be $1.4M_{\odot}$), and $P_{orb}$ is the orbital period (assumed here to be the value given in Equation 1 of 90.53d).

Using these equations and assuming a circular orbit for the neutron star results in $R_{cs} = 1.51 \times 10^{11}$m and $a = 1.54 \times 10^{11}$m. Thus, whilst in outburst, the circumstellar disc does, indeed extend to reach the orbit size supporting the suggestion that the neutron star is constraining the disc from any further growth. 

This also confirms that there could be direct physical interaction between the neutron star and the disc during an X-ray outburst, with the surface area of the disc being temporarily enhanced. The addition of a modest amount of ellipticity in the orbit would further exaggerate this, with the peak interaction occurring around periastron. Hence this offers an explanation why the OGLE I-band magnitude is modulated at the orbital period, and also that the modulated profile seen at this time (see lowest panel in Figure~\ref{fig:folds_all}) tends to exhibit a much broader peak than during times of optical (and X-ray) quiescence (middle panel).

Assuming all the OGLE I-band changes are a direct indication of changes in the circumstellar disc, then the profile of the OGLE data (see Fig.~\ref{fig:ogle}) suggests that the disc had been building up over, at least, the last 2-3 years. It was therefore primed to start feeding material on to the neutron star at the next opportunity - at periastron. So it is perhaps not surprising that the two known X-ray outbursts have both begun around the time of periastron passage. Not only was the OGLE disc at the maximum than it had been for quite a few years, but the extra extension of the disc at periastron suggested by the I band modulation (Fig.~\ref{fig:folds_all}) would have helped bridge the gap and support the onset of accretion.

\subsection{Colour-magnitude variations }

\begin{figure}
	\includegraphics[width=8cm,angle=-0]{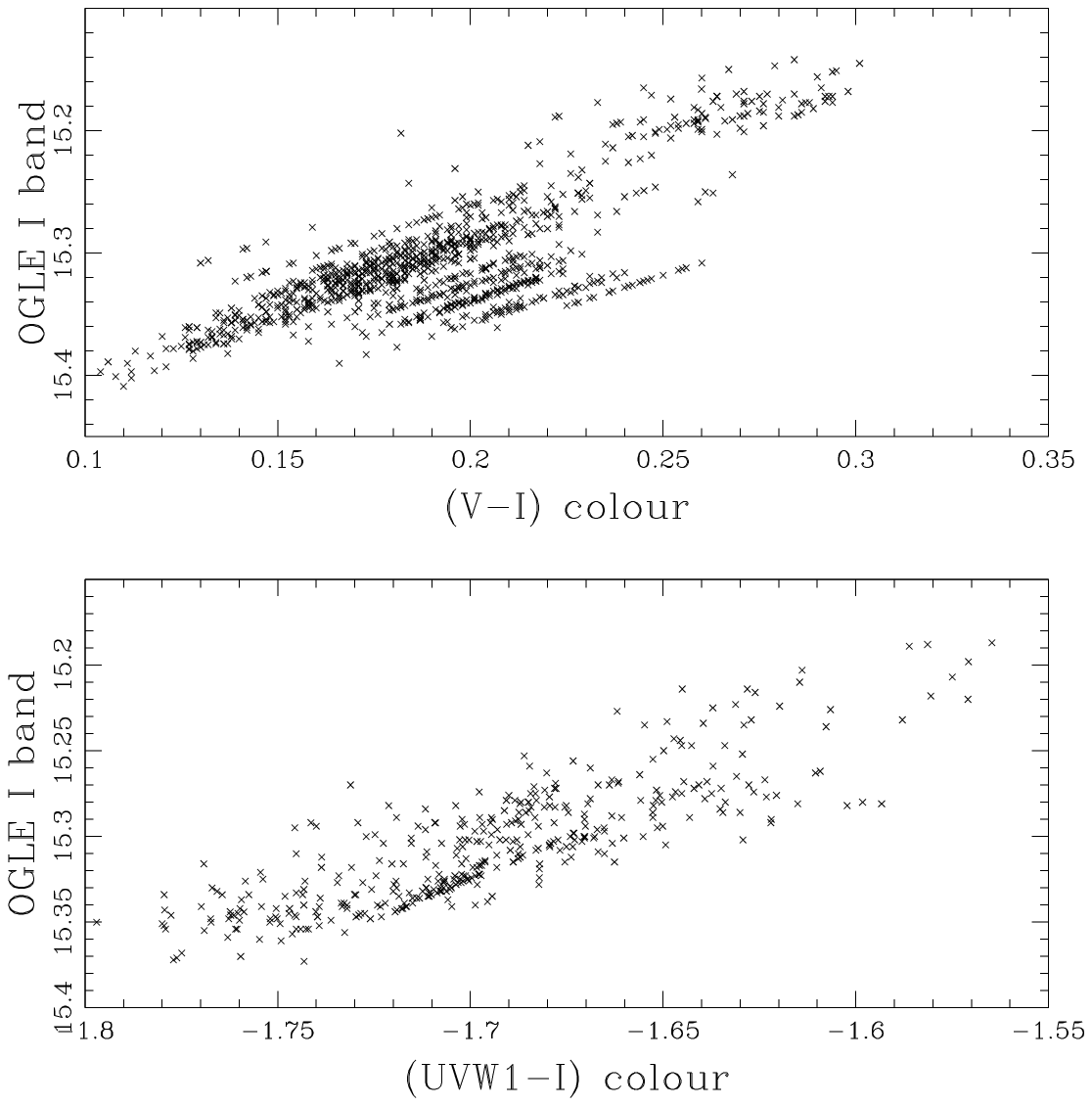}
    \caption{Top panel : {$\sim$25 years worth of OGLE I and V band data used to create a CMD for the behaviour of \scbe{}.}. Lower panel : $\sim$9 years worth of OGLE I band and Swift UVW1 data used to create a similar CMD.}
    \label{fig:cmd3}
\end{figure}

By combining the OGLE I-band data with OGLE V band data it is possible to create a colour-magnitude diagram (CMD) which reveals how the overall colour of the system changed over time and with system brightness. 
The $\sim$24 years worth of OGLE data are used to create a CMD. The data points shown in  the upper panel of Fig.~\ref{fig:cmd3} are calculated by using occasions when both I band and V band measurements were taken within 2 days of each other. 
The same approach may be used by combining the UVOT data with the I band results. Though this combination covers a shorter period of time ($\sim$9 years) it gives a wider spectral coverage and is shown in the lower panel of Fig.~\ref{fig:cmd3}.

The result from both these CMDs is similar - they show a system that reddens as it brightens indicating the growth of a circumstellar disc which has a temperature profile generally cooler than that of the central OB star. This rate of change in colour with a change in I band brightness is approximately twice as large in the (UVW1-I) plot at $\sim$1.5 mag/mag, than in the (V-I) plot of $\sim$0.8 mag/mag. A look at Fig.~\ref{fig:uvot_sed} helps understand why there is this large difference. It is clear that the UVW1 band sits very close to the spectral model profile and hence is a good indicator of the underlying star without a significant circumstellar disc contribution. On the other hand, the V band clearly lies above the model fit indicating that $\sim$25\% of the emission in that band is coming from the circumstellar disc. The I band is even more so, at about $\sim$75\%. So as the disc grows the difference between UVW1 and the I band is always going to be more obvious than between V and I bands.

This pattern of behaviour suggests that we are viewing the circumstellar disc at a low or intermediate inclination angle because an edge-on view would probably produce the opposite effect - fainter when redder. That would happen because a growing circumstellar disc in such a configuration would be expected to increasingly obscure light from the central OB star itself. Nonetheless, the inclination angle does not seem to be large enough to promote a substantial double peaked structure expected in the H$\alpha$ spectral profiles.

\subsection{Imaging of the region surrounding \scbe{}}

The imaging results from the SALTICAM and the SMI200 multi-fibre observations have been presented in Section 2.4 above. It is immediately clear from the H$\alpha$ image (see Fig.~\ref{fig:SALTICAM}) that the extended halo reported by \citep{coe2003} is very prevalent and dominates the region to the SW of \scbe{}. The possible nature of this region has been discussed in detail by those authors with suggestions being SNR, bowshock and HII region. To explore these possibilities the SMI200 observations were carried out permitting a deeper look into the detailed spectral features at various positions around \scbe{}.

A reliable method for distinguishing SNRs and bow shocks from HII regions involves comparing the strengths of the [SII] and H$\alpha$ emission lines. In SNRs and bow shocks, forbidden lines like [SII] are enhanced due to shock excitation, while in HII regions, these lines are less pronounced because the nebulae are photoionized. A high ratio of [SII]/H$\alpha$, typically greater than 0.4, indicates emission from a SNR or bow shock, whereas values below 0.3 suggest emission from an HII region \citep{Fessen1985,Blair2004}. 

\begin{figure}
    \centering

    \begin{subfigure}[b]{0.45\textwidth}
        \centering
        \includegraphics[width=\textwidth]{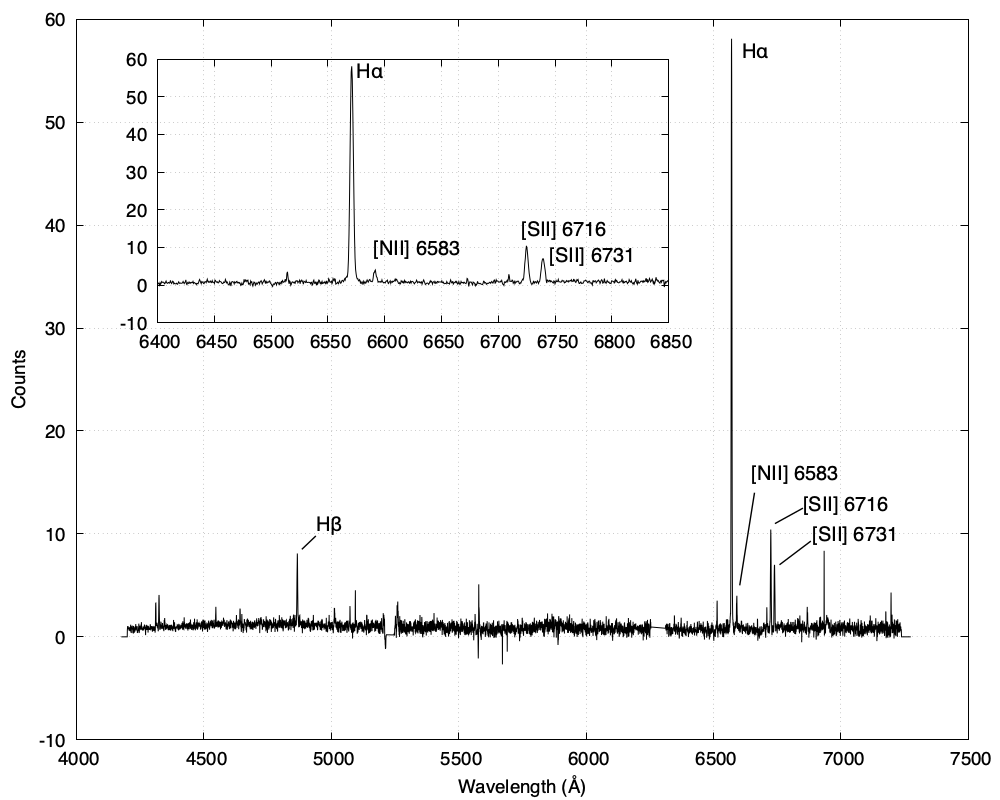}
        \caption{}
        \label{fig:sw_spec}
    \end{subfigure}
    \hfill
    \begin{subfigure}[b]{0.45\textwidth}
        \centering
        \includegraphics[width=\textwidth]{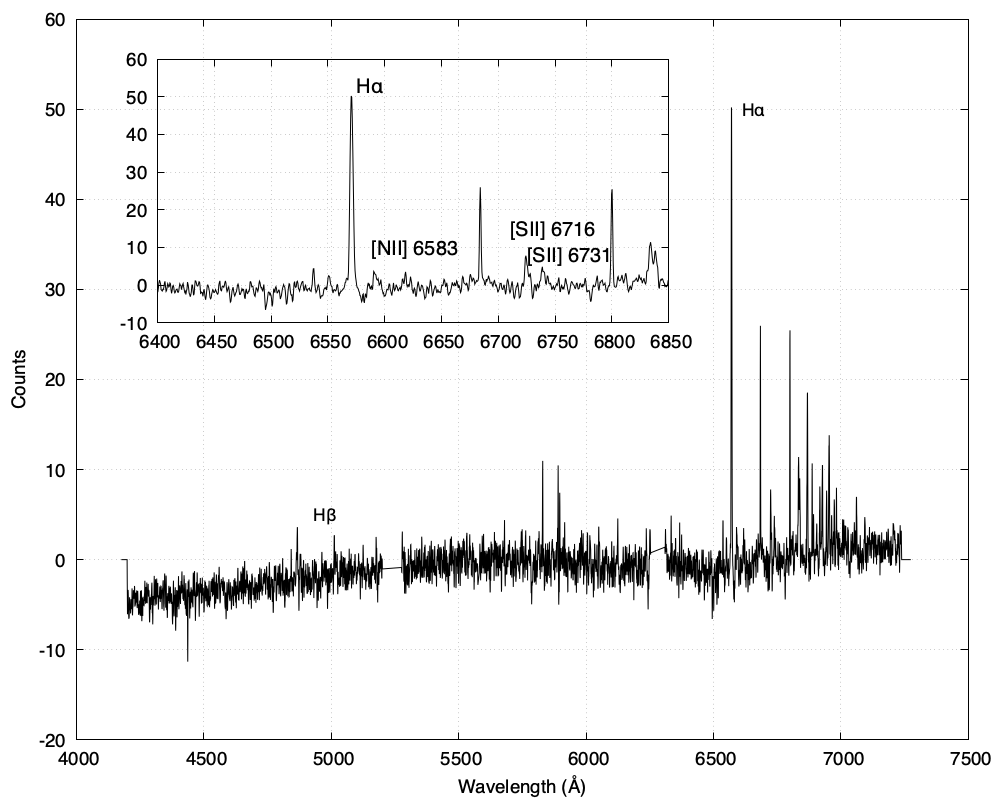}
        \caption{}
        \label{fig:ne_spec}
    \end{subfigure}

    \caption{The combined average IFU spectra from the surrounding region around \scbe{} in the south-west (a) and north-east (b) of Fig.~\ref{fig:IFU}. Each plot contains an insert that zooms in on the wavelength region around H$\alpha$ and the [SII] doublet.}
    \label{fig:IFU_spec}
\end{figure}

To investigate the nature of the environment surrounding the target star, we combined several spectra from the area. Fig.~\ref{fig:IFU_spec} (a) and ~\ref{fig:IFU_spec} (b) show the resultant spectra from the extended emission in the south-west and north-east regions of Fig.~\ref{fig:IFU}, respectively. In both spectra, the [SII]/H$\alpha$ ratio falls within the range of 0.2 to 0.3, which strongly indicates that the extended emission around \scbe{} is due to an HII region rather than a SNR or bow shock.

\section{Conclusions}

In this work, an exceptional episode of X-ray and optical activity by \scbe{} has been observed and investigated. Exceptional not just for a source that has been X-ray quiescent for 26 years, but compared to the general behaviour of the whole cohort of \bexrbs ~systems. For the initial Type II outburst to be immediately followed by a second similar outburst is very unusual and indicative of multiple epochs of material outflow from the mass donor Be star. This is supported by the contemporaneous OGLE data which show that the source has not been this optically bright for over $\sim$16 years. 

Not only is \scbe{} unusual for its current behaviour, but it is also unique in being surrounded by a prominent H$\alpha$ halo. In this paper we have presented the first IFU measurements of this halo and concluded that this phenomenon appears to be one of nature's coincidences. Despite the suggested SNR or bow shock profile of the emission, the system is, by chance, simply surrounded by a HII region. Thus it is thought that this region plays no part in the otherwise exceptional behaviour of \scbe{}.

\section*{Acknowledgements}
A part of this work is based on observations made with the Southern African Large Telescope (SALT), with the Large Science Programme on transients 2021-2-LSP-001 (PI: DAHB). 
The OGLE project has received funding from the Polish National Science
Centre grant OPUS-28 2024/55/B/ST9/00447 to AU.

This work made use of data supplied by the UK Swift Science Data Centre at the University of Leicester. J.A.K. and T.M.G. acknowledge the support of NASA contract NAS5-00136. We acknowledge the use of public data from the Swift data archive.

\section*{Data Availability}

The data underlying this article will be shared on reasonable request to the corresponding author.



\bibliographystyle{mnras}
\bibliography{sc2008} 








\bsp	
\label{lastpage}
\end{document}